\pdfoutput=1
\documentclass[aps,prc,reprint,nofootinbib,showkeys]{revtex4-1}
\usepackage{graphicx}
\usepackage[utf8]{inputenc}
\usepackage{dcolumn}
\usepackage{bm}
\usepackage{slashed}
\usepackage{amsmath,graphicx}
\usepackage[colorlinks=true,linkcolor=blue,citecolor=blue, urlcolor=blue]{hyperref}   
\usepackage{float}
\usepackage{nicefrac}
\usepackage[normalem]{ulem}
\usepackage{amsmath}
\usepackage{subfigure} 
\usepackage{bbold} 
\usepackage[makeroom]{cancel}

\def\be{\begin{equation}}
	\def\ee{\end{equation}}
\newcommand{\bel}[1]{\begin{eqnarray}\label{#1}}
	\newcommand{\eel}{\end{eqnarray}}
\def\barr{\begin{array}}
	\def\earr{\end{array}}
\def\beq{\begin{eqnarray}}
	\def\eeq{\end{eqnarray}}
\def\bfig{\begin{figure}}
	\def\efig{\end{figure}}
\def\lt{\left}
\def\rt{\right}
\newcommand{\nn}{\nonumber}
\newcommand{\f}[2]{\frac{#1}{#2}}

\newcommand{\p}{\partial}

\newcommand{\rf}[1]{Eq.~(\ref{#1})}

\newcommand{\rfn}[1]{(\ref{#1})}

\def\a{\alpha}
\def\b{\beta}

\def\LR{\left(} 
\def\RR{\right)}

\newcommand{\sh}[1]{\sinh#1}
\newcommand{\ch}[1]{\cosh#1}

\newcommand{\lab}[1]{\label{#1}}
\def\nn{\nonumber}

\newcommand{\refeq}[1]{Eq.~(\ref{#1})}

\def\av{{\boldsymbol a}}
\def\bv{{\boldsymbol b}}

\begin{document}
 
    \title{Longitudinal spin polarization in a thermal model} 
	\author{Wojciech Florkowski}
	\email{wojciech.florkowski@uj.edu.pl}
	 \affiliation{M. Smoluchowski Institute of Physics, Jagiellonian University,  PL-30-348 Krak\'ow, Poland}
	\author{Avdhesh Kumar} 
	\email{avdhesh.kumar@ifj.edu.pl} 
	\affiliation{Institute of Nuclear Physics Polish Academy of Sciences, PL-31-342 Krak\'ow, Poland}
\author{Aleksas Mazeliauskas}
\email{a.mazeliauskas@thphys.uni-heidelberg.de}
\affiliation{Institut f\"{u}r Theoretische Physik, Universit\"{a}t Heidelberg, 
69120 Heidelberg, Germany}
	\author{Radoslaw Ryblewski} 
	\email{radoslaw.ryblewski@ifj.edu.pl}
	\affiliation{Institute of Nuclear Physics Polish Academy of Sciences, PL-31-342 Krak\'ow, Poland}
	\date{\today} 
	\bigskip
	\begin{abstract}
  We use a thermal model with single freeze-out to determine longitudinal polarization of $\Lambda$ hyperons emitted from a hot and rotating hadronic medium. We consider the top RHIC energies and use the model parameters determined in the previous analyses of particle spectra and elliptic flow.  Using a direct connection between the spin polarization tensor and thermal vorticity, we reproduce earlier results which indicate a quadrupole structure of the longitudinal component of the polarization three-vector with an opposite sign compared to that found in the experiment. We further use only the spatial components of the thermal vorticity in the laboratory system to define polarization and show that this leads to the correct sign and magnitude of the quadrupole structure. This procedure resembles a non-relativistic connection between the polarization three-vector and vorticity employed in other works. In general, our results bring further evidence that the spin polarization dynamics in heavy-ion collisions may be not directly related to the thermal vorticity. The additional material explains the construction of the hydrodynamicaly consistent gradients of fluid velocity and temperature in thermal models with the help of the perfect-fluid equations of
  motion.

	\end{abstract}
     
\date{\today}

	
	
\maketitle
%
%
\section{Introduction}
 
 Non-central heavy-ion collisions at the relativistic beam energies bring large orbital angular momentum into produced systems. A non-negligible part of such an angular momentum can be further transformed from the initial purely orbital form into the spin part. The latter can be naturally revealed in the spin polarization of emitted particles~\cite{Voloshin:2017kqp,Voloshin:2004ha}. 
 
 Indeed, the spin polarization of $\Lambda$ and  $\bar{\Lambda}$ hyperons has been measured recently by the STAR Collaboration at RHIC~\cite{STAR:2017ckg, Adam:2018ivw}. The result indicates global spin polarization along the direction perpendicular to the reaction plane, which suggests possible connections to the Einstein~--~de~Haas and Barnett effects~\cite{dehaas:1915,RevModPhys.7.129}. 
 
 The experimental results on the global polarization can be successfully explained by the hydrodynamic models~\cite{Karpenko:2016jyx,Li:2017slc}. The basic quantity giving rise to spin polarization in this case is thermal vorticity $\varpi_{\mu \nu}$ defined by the expression $\varpi_{\mu \nu} = -\frac{1}{2} (\p_\mu \b_\nu-\p_\nu \beta_\mu)$, where $\beta_\mu$ is the ratio of the flow velocity $u_\mu$ to local temperature $T$, $\beta_\mu = u_\mu/T$~\cite{Becattini:2007nd,Becattini:2009wh}. The general physics situation is obscured, however, by the fact that the theoretically predicted longitudinal polarization of $\Lambda$'s~\cite{Becattini:2017gcx} has opposite dependence on the azimuthal angle of the emitted particles, as compared to the experimentally found values~\cite{Niida:2018hfw}. 
 
 For our further considerations, it is useful to notice that most of the
 theoretical frameworks used to describe spin polarization deal with particles
 at freeze-out~\cite{Becattini:2013vja,Becattini:2016gvu}. This is natural in
 the approaches that directly connect spin polarization with the ``vortical''
 properties of the fluid~\cite{Karpenko:2016jyx,Xie:2015xpa,Boldizsar:2018akg}.
 In such a scenario, the spin polarization tensor $\omega_{\mu\nu}$ follows
 immediately the space-time changes of the thermal vorticity $\varpi_{\mu
 \nu}$, hence, it is enough to consider the two quantities at freeze-out.
 
 In order to get more insight into the role played by the thermal vorticity at freeze-out, in this work we use one of the versions of the thermal models to analyze the origin of the final longitudinal spin polarization. Thermal models describe very well last stages of heavy-ion collisions (for example, see Refs.~\cite{Cleymans:1992zc,BraunMunzinger:2001ip,Florkowski:2001fp,Becattini:2005xt,Andronic:2017pug}), therefore, they seem to be a natural framework to study the spin polarization of the emitted hadrons such as the $\Lambda$ hyperons. Herein, we use the single freeze-out (SF) model \cite{Broniowski:2001we} which neglect hadronic rescattering in the final state. This model was very successfully used in the past to describe various features of soft hadron production. In particular, it was used for Au+Au collisions at the highest RHIC energies, where the data describing longitudinal spin polarization are now available. Consequently, in this work we do not have to introduce any new parameters --- we rely on the previous estimates. 
In order to calculate the thermal vorticity $\varpi_{\mu\nu}$, we need the knowledge of fluid field gradients on the freeze-out hypersurface. Using perfect-fluid hydrodynamic equations of motion, we derive a general formula for the out-of-plane gradients in terms of the in-plane gradients, which can be calculated directly from the freeze-out surface parametrization.

 Our first results reported below, based on the tight connection of the spin polarization tensor $\omega_{\mu\nu}$ with the thermal vorticity $\varpi_{\mu\nu}$, confirm that the longitudinal spin polarization has a~quadrupole structure with an opposite sign compared to the measured signal. To study the spin polarization effects in more detail we explore yet another case, where the spin polarization tensor is not directly related to the thermal vorticity.
 
 The idea that the spin polarization tensor can evolve independently from the thermal vorticity was put forward first in Ref.~\cite{Florkowski:2017ruc} and developed in Refs.~\cite{Florkowski:2017dyn,Florkowski:2018myy,Florkowski:2018ahw,Becattini:2018duy} (for a recent review see Ref.~\cite{Florkowski:2018fap} and for related works see Refs.~\cite{Sun:2018bjl,Montenegro:2018bcf,Weickgenannt:2019dks,Hattori:2019lfp,Xie:2019jun}).  In the perfect-fluid approach to hydrodynamics with spin, proposed in Ref.~\cite{Florkowski:2017ruc}, the space-time evolution of the spin polarization tensor is determined by the conservation law for the total angular momentum (we note that for particles with spin this conservation law takes a non-trivial form). An example of such an evolution, in the case of a simple one-dimensional and boost-invariant expansion, has been analyzed recently in Ref.~\cite{Florkowski:2019qdp}.

The model used in this work is also boost-invariant but includes a non-trivial transverse hydrodynamic expansion that leads to vortical structures in the transverse plane that, in turn, can induce the longitudinal spin polarization at midrapidity, if a certain type of relation between vorticity and spin polarization is assumed. However, due to the assumed boost invariance, the present approach yields zero polarization in the transverse direction at midrapidity.

Besides the case where the spin polarization tensor is directly defined in terms of the thermal vorticity, $\omega_{\mu\nu} = \varpi_{\mu \nu}$, we also consider the case where only the spatial components of the thermal vorticity in the laboratory (LAB) frame are taken into account. In the latter case, dubbed below as the case with the projected thermal vorticity, we assume that $\omega_{\mu\nu} = \varpi_{\alpha \beta} \bar\Delta^\alpha_{\,\,\mu} \bar\Delta^\beta_{\,\,\nu}$, where $\bar \Delta^{\mu\nu} = g^{\mu \nu}-u_{\rm LAB}^\mu u_{\rm LAB}^\nu $, $u_{\rm LAB}^\mu = (1,0,0,0)$ and the metric tensor is chosen as  $g_{\mu\nu} =  \hbox{diag}(+1,-1,-1,-1)$. Such a relation is similar to the non-relativistic treatment of the polarization-vorticity coupling, which is able to correctly describe the sign of the longitudinal polarization~\cite{Voloshin}. Indeed, it turns out that with the choice $\omega_{\mu\nu} = \varpi_{\alpha \beta} \bar\Delta^\alpha_{\,\,\mu} \bar\Delta^\beta_{\,\,\nu}$ one can describe the quadrupole structure of the longitudinal polarization with the correct sign. As a consequence, our results give further evidence  that the dynamics of spin polarization may be decoupled from the space-time behavior of the thermal vorticity.

{\it Notation and conventions:} Unless specified otherwise, the scalar product of two four-vectors $a^{\mu}$ and $b^{\mu}$ is denoted by $a \cdot b =a^{\mu}b_{\mu}= g_{\mu \nu} a^\mu b^\nu = a^0 b^0 - \av \cdot \bv$, where bold font is used to represent three-vectors. The convention $\epsilon^{0123} = -\epsilon_{0123} =+1$ is used for the Levi-Civita tensor $\epsilon^{\mu\nu\rho\sigma}$.  Natural units $\hbar = c= k_B~=1$ are used throughout the text.
%
\section{\label{sec:sf} Single freeze-out model}
%
\subsection{\label{sec:sf0} General concept}

In the thermal SF model one assumes that the chemical and thermal freeze-outs coincide, i.e., there is no hadronic rescattering included after the chemical freeze-out. The chemical freeze-out is assumed to take place on a space-time hypersurface where all hadrons (stable and unstable with respect to strong interactions) are created. Unstable hadrons decay, giving contributions to the yields of stable hadrons. At this level one can perform a traditional analysis of the ratios of hadronic abundances and determine thermodynamic parameters characterizing the chemical freeze-out, such as the freeze-out temperature $T$ and baryon chemical potential $\mu$. 

Besides the ratios of hadronic yields, the assumption about the single freeze-out allows us to directly calculate hadronic spectra --- provided one knows the hydrodynamic flow of matter on the freeze-out hypersurface, $u^\mu$, as well as the space-time geometry of the freeze-out hypersurface. As a matter of fact, the form of the flow can be treated as a model input, to be determined from the analyses of the spectra of various particles. We note that for boost-invariant systems, one fits only the transverse-momentum spectra. 

In its original formulation, aiming at the description of heavy-ion collisions at very high energies, the SF model has four parameters: two thermodynamic ones and two geometric ones. The two thermodynamic parameters, temperature $T$ and baryon chemical potential $\mu$, are fitted from the ratios of hadronic abundances. The two geometric parameters, $\tau_f$ and $r_{\rm max}$, characterize the freeze-out hypersurface and the hydrodynamic flow. The freeze-out hypersurface is defined by the conditions: $\tau^2_f = t^2 - x^2 - y^2 - z^2$ and $x^2 + y^2 \leq r^2_{\rm max}$. The hydrodynamic flow has the Hubble-like form, $u^\mu = x^\mu/\tau$. 

\subsection{\label{sec:sf1} Asymmetry in transverse plane}

In order to include the phenomena such as an elliptic flow, the original version of the SF model was extended to include the elliptic deformations of both the emission region in the transverse plane and of the transverse flow~\cite{Broniowski:2002wp}. This was achieved by using the following parametrization of the boundary region in the transverse plane,
\begin{align}
x &= r_{\rm max} \sqrt{1-\epsilon} \cos\phi, \nn \\
y &= r_{\rm max} \sqrt{1+\epsilon} \sin\phi
.\end{align}
Here $\phi$ is the azimuthal angle, while $r_{\rm max}$ and $\epsilon$ are the model parameters. With $\epsilon >0$ the system formed in the collisions is elongated in the $y$ direction, i.e., out of the reaction plane (resembling a characteristic almond shape).

The asymmetric flow profile is accordingly defined as follows
\begin{align}
u^\mu&=\LR\frac{t}{N},   \frac{x}{N} \sqrt{1+\delta}, \frac{y}{N} \sqrt{1-\delta },   \frac{z}{N}\RR,\label{eq:u}
\end{align}
where $\delta$ is a parameter accounting for the transverse flow anisotropy.
For $\delta > 0$, there is more flow in the reaction plane, an effect that can be identified as the elliptic flow. Using the normalization condition $u^{\mu}u_{\mu}=1$, one can determine the normalization factor in~Eq.~(\ref{eq:u}), 
\beq
N={\sqrt{\tau ^2-\left(x^2-y^2\right)\delta}}\,,
\label{eq:N}
\eeq
where $\tau$ is the proper time
\beq
\tau^2 = t^2 -x^2-y^2-z^2. 
\label{eq:tau}
\eeq
The parameters $\epsilon$ and $\delta$ are two additional parameters needed to describe the effects of non-trivial dynamics in the transverse plane in the case of non-central collisions. We note that all our parametrizations hold in the LAB frame which can be identified with the center-of-mass frame of the colliding nuclei.

In the following, we assume that freeze-out takes place at a constant value of the proper time, temperature and chemical potential, i.e., at $\tau = \tau_f$, $T=T_{f}$ and $\mu=\mu_f$.  In this case a three-dimensional element of the freeze-out hypersurface, $\Delta\Sigma_\lambda$, is given by the formula
\beq
\Delta \Sigma_{\lambda } &=& n_{\lambda }\, dx dy\, \tau_f d\eta \lab{sig}, 
\eeq
where the surface norm vector is given by
\beq
n^{\lambda }\!=\!\frac{1}{\tau_f} \left(\sqrt{\tau^2_f\!+\!x^2\!+\!y^2}\ch\eta,x,y,\sqrt{\tau^2_f\!+\!x^2\!+\!y^2}\sh\eta \right).\nn\\
\label{eq:n}
\eeq
Here $\eta=\frac{1}{2} \ln \left[(t+z)/(t-z)\right]$ is the space-time rapidity. One can easily notice that \mbox{$n^\lambda n_\lambda = +1$}.
\subsection{\label{sec:thvort} Thermal vorticity}

The thermal vorticity defined above can be rewritten as a sum of the two terms,
\beq
\varpi _{\mu \nu }=-\frac{1}{2T}\lt(\p_{\mu}u_{\nu}-\p_{\nu}u_{\mu}\rt) - \frac{1}{2T^2}\lt( u_{\mu} \partial_\nu T - u_{\nu} \partial_\mu T \rt). \nonumber \\
\label{eq:thvor}
\eeq
The parametrization of the hydrodynamic flow introduced above allows us to determine the first term on the right-hand side of Eq.~(\ref{eq:thvor}). However, the second term contains temperature gradients that are not defined in the thermal model --- we only know that the temperature gradients calculated in directions that are parallel to the freeze-out hypersurface should vanish. This problem can be overcome by assuming that the temperature gradients follow from the hydrodynamic calculations. In the physics case discussed herein, we can assume that the baryon number density can be to a first approximation neglected and use the hydrodynamic equations summarized in Appendix~\ref{sec:hydmub0}. A direct calculation that uses Eq.~(\ref{eq:tempgrad}) shows that the second term on the right-hand side of Eq.~(\ref{eq:thvor}) is exactly equal to the first term\footnote{As shown in Appendix~\ref{sec:hydmub0}, this property is independent of the form of sound velocity (including its temperature dependence), but may depend on our choice of the flow at freeze-out. It may also change if dissipative hydrodynamics is used instead of the perfect-fluid approached employed here. Interestingly, the numerical calculations presented in \cite{Karpenko:2018erl} show a similar pattern to our model calculations.}. Using this fact we obtain all the components of $\varpi _{\mu \nu }$: 

\beq
\varpi _{01}&=&\frac{t x}{T N^{3}}  \left(1+\delta-\sqrt{1+\delta}\right), \nn \\
\varpi _{02}&=&-\frac{t y}{T N^{3}} \left(\sqrt{1-\delta }-1+\delta\right), \nn \\
\varpi _{03}&=&0, \nn \\
\varpi _{12}&=&\frac{x y \sqrt{1-\delta ^2}}{T N^{3}}  \left(\sqrt{1+\delta}-\sqrt{1-\delta }\right), \nn \\
\varpi _{23}&=&-\frac{y z}{T N^{3}} \left(\sqrt{1-\delta }-1+\delta \right), \nn \\
\varpi _{13}&=&\frac{x z}{T N^{3}} \left(1+\delta-\sqrt{1+\delta}\right).
\label{eq:thvor1}
\eeq

It is important to emphasize at this point that the temperature gradients determined from the hydrodynamic flow might be inconsistent with the form of the freeze-out hypersurface. A constant temperature $T$ on the freeze-out hypersurface $\Sigma$ requires that the gradient $\partial_\mu T$ is proportional to the four-vector $n_\mu$ defined by Eq.~(\ref{eq:n}). In our case, we have checked that this condition holds if one keeps the terms linear in $\delta$ only. Since in the numerical calculations we use small values of $\delta$, our treatment of freeze-out is to a very good approximation consistent with perfect-fluid hydrodynamic description.

In a general case, the hydrodynamic description of a fluid uses space-like gradients of
hydrodynamic quantities to determine the time-like gradients.
This allows to solve the full space-time evolution of a system from initial conditions.
Although a SF model does not specify 
the entire pre-history of such hydrodynamic evolution,
we can nevertheless use the hydrodynamic equations of motion to self-consistently
determine the  gradients of fluid fields. 
In
Appendix~\ref{sec:scg}, we show how the gradients of hydrodynamic variables on a constant temperature $T$ and chemical potential  $\mu$ freeze-out surface (in-plane derivatives), can
be used to directly determine the orthogonal gradients (out-of-plane
derivatives). These results might be useful in the formulations of thermal
models that are automatically consistent with perfect-fluid hydrodynamics. 

\subsection{\label{sec:modpar} Model parameters}

In the form defined above, the SF model has altogether six parameters: $T=T_f$ (freeze-out temperature), $\mu$ (freeze-out baryon chemical potential), $r_{\rm max}$ (transverse size), $\tau_f$ (system's lifetime), $\epsilon$ (size deformation) and $\delta$ (flow deformation). The first two are determined solely by the ratios of hadronic yields. In the case numerically studied below we use $T_f=$~165 MeV. The value of the baryon chemical potential is irrelevant for our study, since it cancels in the expressions for the mean polarization (a consequence of the use of classical statistics that is appropriate for heavy particles such as $\Lambda$'s).

The remaining four parameters should be obtained from the fits of the hadron spectra and elliptic flow. The analyses of this type were performed in the past and the resulting values of the parameters describing various reactions studied in different centrality bins can be found in \cite{Florkowski:2004du}. The values used herein are listed in Table~\ref{tab}.

\section{\label{sec:level2} Spin polarization of particles}
%
\subsection{Pauli-Luba\'nski (PL) four-vector}
The mean spin polarization of particles can be directly obtained from the Pauli-Luba\'nski (PL) vector. The latter is first calculated in the LAB frame for particles that are produced on the freeze-out hypersurface with momentum $p$. Subsequently, it is boosted to the rest frame of those particles. In this frame, the PL vector has only space-like components. We divide them by the number of particles with momentum $p$ to get the mean polarization. The mean spin polarization is a three-vector whose components depend on the momenta of particles. At midrapidity, the longitudinal polarization can be studied as a function of transverse-momentum components $p_x$ and $p_y$.

 The phase-space density of the PL four-vector $\Pi_{\mu}$ is given by the expression~\cite{Florkowski:2017dyn}
\begin{equation}
E_p\frac{d\Delta \Pi _{\mu }(x,p)}{d^3 p}
=-\frac{1}{2}\epsilon _{\mu \nu \alpha \beta }\Delta 
\Sigma _{\lambda }E_p\frac{dS^{\lambda ,\nu \alpha }_{\rm GLW}(x,p)}
{d^3 p}\frac{p^{\beta }}{m}.
\label{PL1}
\end{equation}
The particle four-momentum $p^\lambda$ can be parametrized in terms of the transverse momentum~$p_T = \sqrt{p_x^2+p_y^2}$ \,, rapidity~$y_p$, and the azimuthal angle $\phi_p$,
\begin{align}
p^\lambda &= \left( E_p,p_x,p_y,p_z \right)\nn \\
          &= \left( m_T\ch y_p,p_T \cos\phi_p,p_T \sin\phi_p,m_T\sh y_p \right). \lab{pl}
\end{align}
Here $m_T = \sqrt{m^2 + p_T^2}$ \, is the transverse mass while $m$ is the mass of the particle. In the numerical calculations we assume that $m$ is equal to the $\Lambda$ hyperon mass. 

The expression $dS^{\lambda ,\nu \alpha }_{\rm GLW}(x,p)/d^3 p$ in \rf{PL1} denotes the phase-space density of the spin tensor obtained in the GLW    kinetic theory framework (de Groot, van~Leeuwen, van~Weert~\cite{DeGroot:1980dk}). It is given by the formula~\cite{DeGroot:1980dk,Florkowski:2018ahw}
\begin{align}
E_p \f{dS^{\lambda , \nu \a }_{\rm GLW}}{d^3p} &=&\frac{{\cosh}(\xi)}{(2\pi)^3 m^2} \, e^{-p \cdot \beta} p^{\lambda } \left(m^2\omega ^{\nu\a}+2 p^{\delta }p^{[\nu }\omega ^{\a ]}{}_{\delta } 
\right), \lab{eq:SGLW22} 
\end{align}
where $\xi$ is the ratio of the (baryon) chemical potential and the temperature, $\xi=\mu/T$. Using \rf{eq:SGLW22} in \rf{PL1} and integrating over the freeze-out hypersurface, we can define the total value of the PL four-vector for particles with momentum $p$,
\begin{equation}
E_p\frac{d\Pi _{\mu }(p)}{d^3 p} = -\f{ \cosh(\xi)}{2 (2 \pi )^3 m}
\int
e^{-\beta \cdot p}\,\Delta \Sigma \cdot p \,
 \epsilon _{\mu \beta \rho \sigma } \omega^{\rho\sigma} p^{\beta }. \lab{PDPLV}
\end{equation}

Now at the freeze-out, we can write
\beq
p \cdot\beta&=&R_1 \cosh(y_p-\eta)+R_2\, ,
\lab{eq:pu} 
\eeq
In the above expression the functions $R_1$ and $R_2$ are defined by
\beq
R_1&=&\frac{m_T \sqrt{\tau_f ^2+x^2+y^2}}{T_f N_f}, 
\nn \\
R_2 &=&-\frac{x p_x \sqrt{1+\delta}+y p_y \sqrt{1-\delta }}{T_f N_f}.\nn \\ \lab{R1R2}
\eeq
where, $T_f$ is the freeze-out temperature and $N_f={\sqrt{\tau_f ^2-\left(x^2-y^2\right)\delta}}$.

In the similar way, we can also write, 
\beq
\Delta \Sigma \cdot p&=&\left[ G_1 \cosh \left(y_p-\eta \right)+G_2\right] dx dy d\eta, 
\lab{SIGP}   
\eeq
where
\beq
G_1&=&m_T \sqrt{\tau_f ^2+x^2+y^2}, \quad
G_2=-(x p_x+y p_y). \lab{G1G2}
\eeq

\subsection{Spin polarization defined by thermal vorticity}

In this section we assume that thermal vorticity is equal to spin polarization. In this case, the contraction of the dual polarization tensor and four-momentum can be written in a compact form as 
\begin{equation}
\frac{1}{2} \epsilon _{\mu \beta \rho \sigma } \varpi^{\rho\sigma}p^{\beta}=\left[
\begin{array}{c}
 G_{00}\sh(\eta)+G_{01}\sh(y_p)  \\
G_{10}\sh(y_p-\eta)  \\
 G_{20} \sh(y_p-\eta)  \\
-G_{00}\ch(\eta)-G_{01}\ch(y_p) \\ 
\end{array}
\right]\,,\lab{OP}
\end{equation}
where we defined the following auxiliary functions:
%
\beq
G_{00}&=&-\frac{\sqrt{\tau_f^2+x^2+y^2}}{T_f N_f^{3}}\Big[ y p_x ((1-\delta)-\sqrt{1-\delta})\nn\\
      &&\qquad\qquad-x p_y ((1+\delta)-\sqrt{1+\delta})\Big], \nn\\
G_{01}&=&-\frac{x y m_T}{T_f N_f^{3}}\sqrt {1-\delta^2}(\sqrt{1+\delta}-\sqrt{1-\delta}), \nn\\
G_{10}&=&-\frac{y m_T}{T_f N_f^{3}} \sqrt{\tau_f^2+x^2+y^2} \lt[ (1-\delta)-\sqrt{1-\delta}\rt], \nn\\
G_{20}&=&\frac{x m_T}{T_f N_f^{3}} \sqrt{\tau_f^2+x^2+y^2} \left[ (1+\delta)-\sqrt{1+\delta} \right].
\eeq
Now using Eqs. (\ref{eq:pu}), (\ref{SIGP}) and (\ref{OP}) in \rf{PDPLV}, the total PL vector can be expressed as
\begin{equation}
E_p\frac{d\Pi _{\mu }(p)}{d^3 p}= \frac{\cosh(\xi)}{(2 \pi )^3 m} \left[\begin{array}{c}
-\sinh \left(y_p\right)\int \limits_{A}e^{-R_2} F_1 dx dy  \\
  0 \\
 0  \\
\cosh \left(y_p\right)\int \limits_{A}e^{-R_2} F_1 dx dy \\
\end{array}
\right],
\label{totPL}
\end{equation}
where
\beq
F_1 &=&2 \left(G_{01} G_1+G_{00} G_2\right) K_1 (R_1)+2 G_{01} G_2 K_0 (R_1)\nn\\
    &&+G_{00} G_1 \left(K_0(R_1)+K_2(R_1)\right).\eeq
Here $K_n$'s  are the modified Bessel functions of the second kind.  

\subsection{\label{sec:5} The mean PL four-vector}
The mean PL four-vector is defined as a ratio of the total PL vector~\rfn{totPL} and the momentum density of all particles (i.e., of both particles and antiparticles)
\beq
\langle\pi_{\mu}\rangle=\frac{E_p\frac{d\Pi _{\mu }(p)}{d^3 p}}{E_p\frac{d{\cal{N}}(p)}{d^3 p}}. \lab{avPL}
\eeq
Here we can use the formula \cite{Florkowski:2019qdp}
\beq
E_p\frac{d{\cal{N}}(p)}{d^3 p}&=&
\f{4 \cosh(\xi)}{(2 \pi )^3}
\int \Delta \Sigma _{\lambda } p^{\lambda } 
\,
e^{-\beta \cdot p} \,. \lab{MD}
\eeq
Substituting Eqs.~\rfn{eq:pu} and \rfn{SIGP} into \rf{MD} we can get
\beq
E_p\frac{d{\cal{N}}(p)}{d^3 p}&=&
 \frac{8 \cosh(\xi)}{(2 \pi )^3} \int\limits _{A} e^{-R_2} F_2 \, dx dy,  \lab{MD1}
\eeq
where
\beq
F_2 = G_1K_1(R_1)+G_2K_0(R_1).
\eeq

 The mean PL four-vector $\langle\pi^{\star}_{\mu}\rangle$ in the particle rest frame can be obtained by using the canonical boost \cite{Leader:2001}
 \beq
 \Lambda(-{\boldsymbol{v}}_p)&=&
 \begin{bmatrix}
 \frac{E_p}{m} & -\frac{p_x}{m} & -\frac{p_y}{m} & -\frac{p_z}{m} \\
 -\frac{p_x}{m} & 1+\alpha _p p_x^2 & \alpha _p p_x p_y & \alpha _p p_x p_z \\
 -\frac{p_y}{m} & \alpha _p p_y p_x & 1+\alpha _p p_y^2 & \alpha _p p_y p_z \\ 
 -\frac{p_z}{m} & \alpha _p p_z p_x & \alpha _p p_z p_y & 1+\alpha _p p_z^2 
\end{bmatrix},\nn\\
\eeq
where $\alpha_p = 1/(m (E_p + m))$. In this way we find
\begin{equation}
\langle\pi^{\star}_{\mu}\rangle=\frac{H}{8m (m_T \cosh y_p + m)}\left[\begin{array}{c}
0 \\ 
 -p_x \sinh y_p \\ 
 -p_y \sinh y_p  \\ 
m \cosh y_p + m_T \\
\end{array}
\right],
\label{meanPL}
\end{equation}
where
\beq 
H= \frac{\int \limits_{A} e^{-R_2}F_1 dx dy}{\int \limits_{A} e^{-R_2} F_2 dx dy}.
\eeq 
It can be easily shown that $\langle\pi^{\star}_{\mu}\rangle\langle\pi_{\star}^{\mu}\rangle=\langle\pi_{\mu}\rangle\langle\pi^{\mu}\rangle=-P^2 = -H^2/(64 m^2)$.

\subsection{The case with projected thermal vorticity}
\label{sec:projectedtv}

In order to check how the polarization effects may depend on the coupling between the spin polarization tensor and the thermal vorticity, we consider herein also the case where the spin polarization tensor is defined by the expression $\omega_{\mu\nu} = \varpi_{\alpha \beta} {\bar \Delta}^\alpha_{\,\,\mu} {\bar \Delta}^\beta_{\,\,\nu}$. Here ${\bar \Delta}^{\mu\nu} =  g^{\mu \nu} -u_{\rm LAB}^\mu u_{\rm LAB}^\nu$ and $u_{\rm LAB}^\mu = (1,0,0,0)$. This choice corresponds to setting $\omega_{i j} = \varpi _{ij}$ and  $\omega_{0i} = 0$ in the previously discussed expressions. A straightforward calculation leads in this case to the formula
\beq 
\frac{1}{2} \epsilon _{\mu \beta \rho \sigma } \omega^{\rho\sigma} p^{\beta }=\left[
\begin{array}{c}
 G_{00}\sh(\eta)+G_{01}\sh(y_p)  \\
-G_{10}\sh(\eta)\ch(y_p)  \\
-G_{20} \sh(\eta)\ch(y_p) \\
-G_{01}\ch(y_p) \\ 
\end{array}
\right].
\lab{OP1}
\eeq 
Using Eqs. (\ref{eq:pu}), (\ref{SIGP}) and (\ref{OP1}) in \rf{PDPLV}, the total PL vector can be expressed as
\begin{equation}
\!E_p\frac{d\Pi _{\mu }(p)}{d^3 p}\!=\!\frac{\cosh(\xi)}{(2 \pi )^3 m} \left[\begin{array}{c}
 -\sinh \left(y_p\right)\!\int\limits _{A}\!e^{-R_2} F_1 dx dy  \\
 \sinh \left(y_p\right)\cosh\left(y_p\right) \!\int\limits _{A}\!e^{-R_2} L_1 dx dy\\
 \sinh \left(y_p\right)\cosh\left(y_p\right) \!\int\limits _{A}\!e^{-R_2} L_2 dx dy  \\
 \cosh \left(y_p\right)\!\int\limits _{A}\!e^{-R_2} L_3 dx dy \\
 \end{array}
 \right],
 \label{totPL1}
 \end{equation}
 where
 \beq
 L_1 &=&G_1 G_{10} (K_0(R_1)+K_2(R_1))+2 G_2 G_{10} K_1 (R_1),\nn\\
 L_2 &=&G_1 G_{20} (K_0(R_1)+K_2(R_1))+2 G_2 G_{20} K_1 (R_1),\nn\\
 L_3 &=&2 G_{01} G_1 K_1 (R_1)+2 G_{01} G_2 K_0 (R_1).
 \eeq
 Using Eqs.~\rfn{MD1} and \rfn{totPL1} in \rf{avPL}, we can calculate the mean PL four-vector. Then, by boosting the mean polarization to the particle rest frame, we find the following formula for the mean longitudinal component of the PL four-vector
\begin{align}
  \langle\pi^{\star}_{z}\rangle=-\frac{1}{8m}\Bigg[&
\frac{(m\cosh y_p\!+\!m_T)}{(m_T \cosh y_p\!+\! m)}\frac{\int\limits _{A}e^{-R_2} L_3 dx dy}{\int\limits _{A} e^{-R_2} F_2 dx dy}\\
                                                  &+\frac{ m_T \sinh^2 y_p}{(m_T \cosh y_p\!+\!m)}\frac{\int\limits _{A}e^{-R_2} (L_3-F_1) dx dy}{\int\limits _{A} e^{-R_2} F_2 dx dy}
\Bigg] \nn 
\label{lgPL} 
.\end{align}
This expression is used in our numerical calculations presented below.

\section{Results and discussions}
 \begin{table*}
\centering
\begin{tabular}{ |p{3cm}||p{3cm}||p{3cm}||p{3cm}||p{3cm}|} 
   \hline
 c $\%$ & $\epsilon$ &   $\delta$ &   $\tau_f$~[fm] &   $r_{\rm max}~[fm]$ \\
  \hline
   $0-15$ &   $0.055$ &   $0.12$ &   $7.666$ &   $6.540$ \\
  $15-30$ &   $0.097$ &   $0.26$ &   $6.258$ &   $5.417$ \\
  $30-60$ &   $0.137$ &   $0.37$ &   $4.266$ &   $3.779$ \\
 \hline
\end{tabular}
\caption{Thermal model parameters used to describe the PHENIX data ($\sqrt{s_{NN}}=130$~GeV), see \cite{baran}.}
\label{tab}
\end{table*}
In this section we present our numerical results for the longitudinal component of the mean PL four-vector, which describes the longitudinal spin polarization of $\Lambda$-- hyperon ($m=1.116$~GeV). The model parameters used in the calculations are given in Table~\ref{tab}. They were fitted before to describe the PHENIX data at the beam energy $\sqrt{s_{NN}}=130$~GeV \cite{baran}, for the three centrality classes: $c$=0--15\%, $c$=15--30\%, and $c$=30--60\% at freeze-out temperature $T_f=0.165$~GeV.

\begin{figure*}
      \centering
		\subfigure[]{}\includegraphics[width=0.48\textwidth]{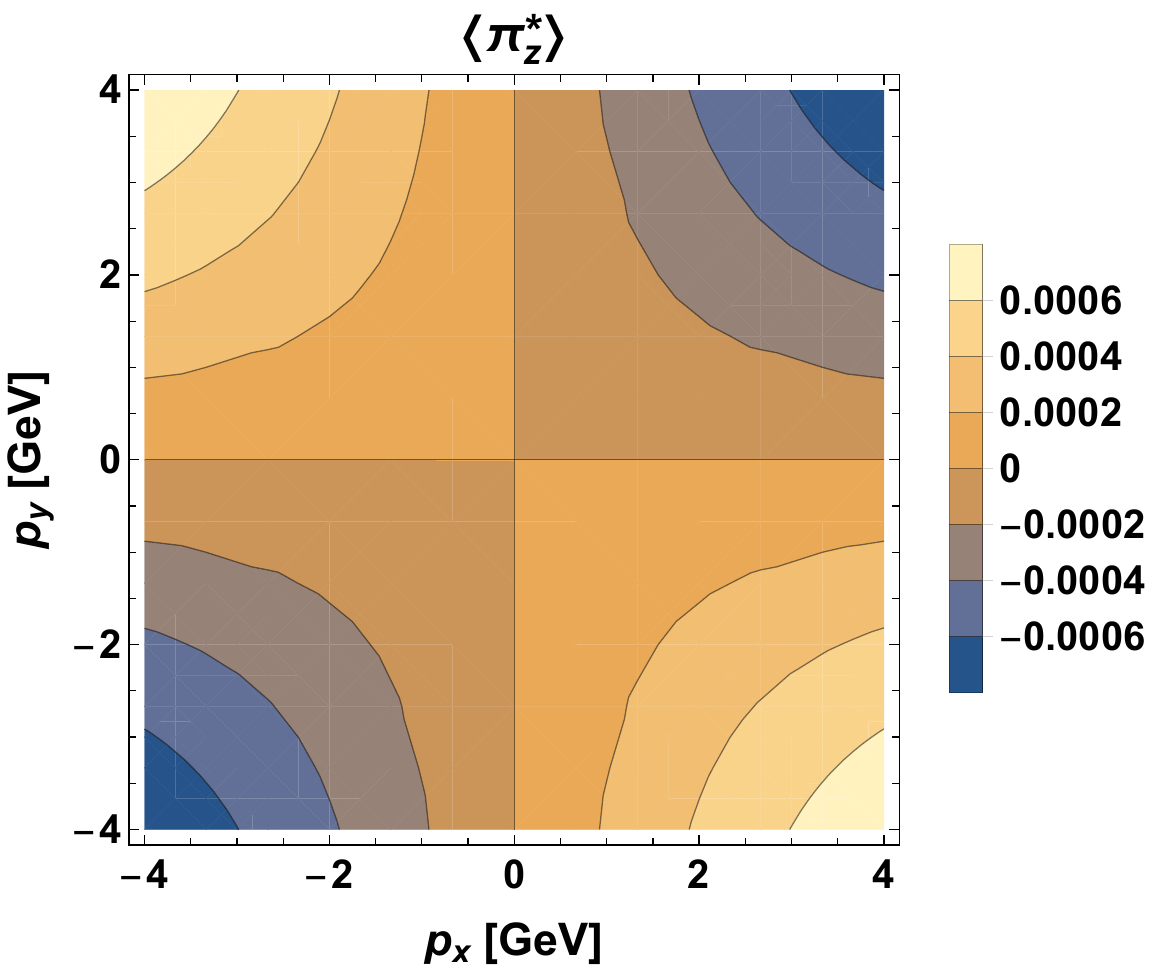}
		\subfigure[]{}\includegraphics[width=0.48\textwidth]{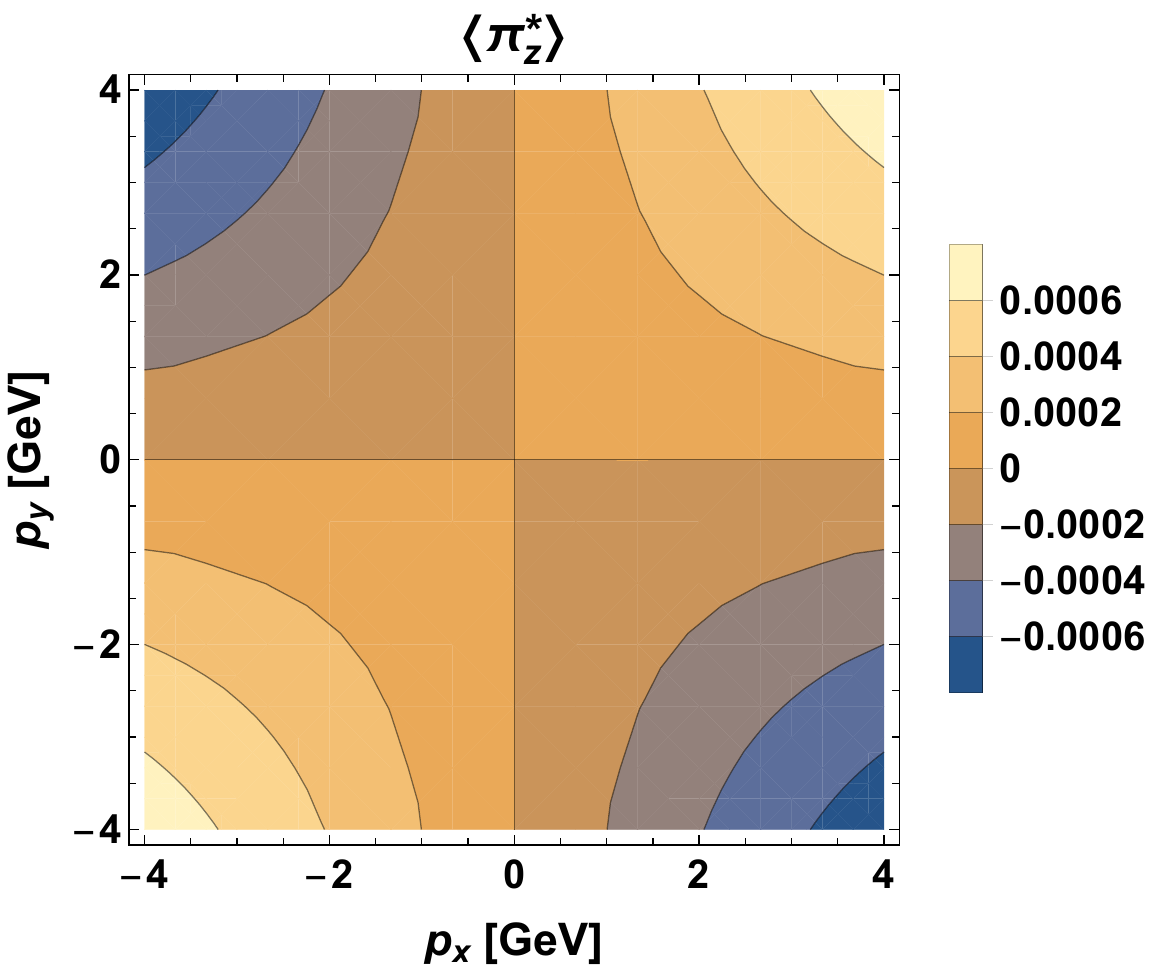}
		\caption{Longitudinal component of the PRF mean polarization three-vector of the $\Lambda$ hyperons for the centrality class $c$=0--15\%.  Panel (a) describes the case where the spin polarization is defined by the thermal vorticity, panel (b) corresponds to the case where we use the projected thermal vorticity defined in Sec.~\ref{sec:projectedtv}.} 
	 \label{fig:polarization1}
\end{figure*}
\begin{figure*}
      \centering
		\subfigure[]{}\includegraphics[width=0.48\textwidth]{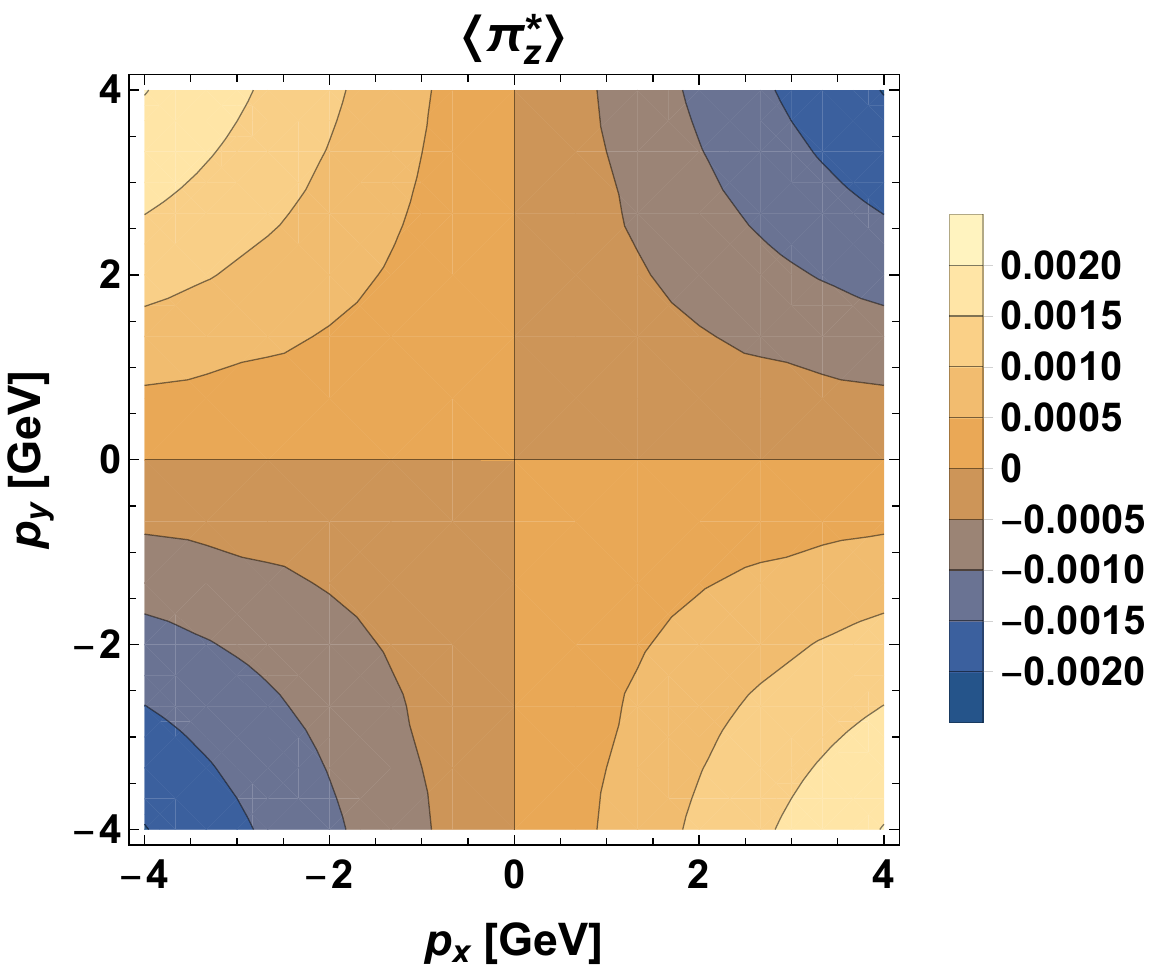}
		\subfigure[]{}\includegraphics[width=0.48\textwidth]{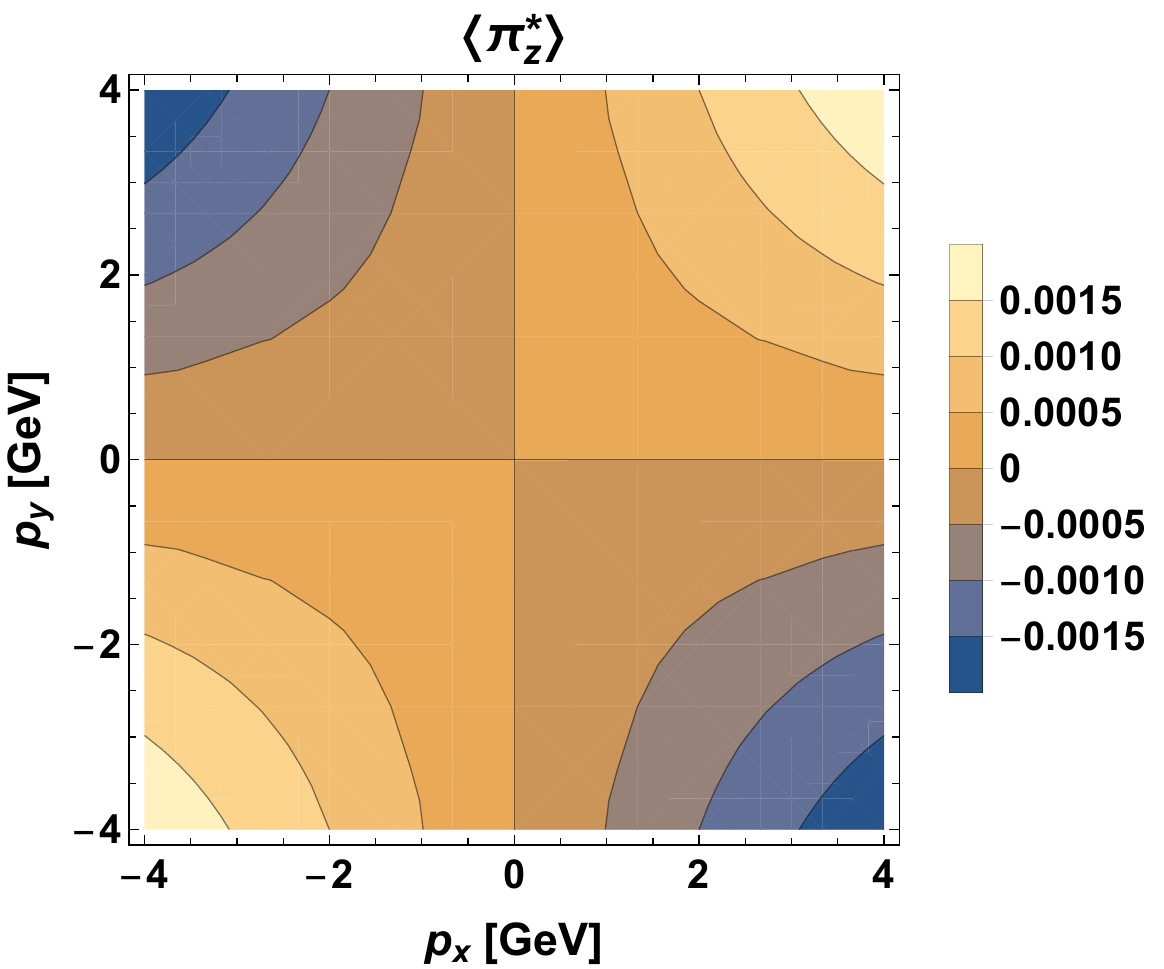}
		\caption{Same as Fig.~\ref{fig:polarization1} but for the centrality class $c$=15--30\%.} 
	 \label{fig:polarization2}
\end{figure*}
\begin{figure*}
      \centering
		\subfigure[]{}\includegraphics[width=0.48\textwidth]{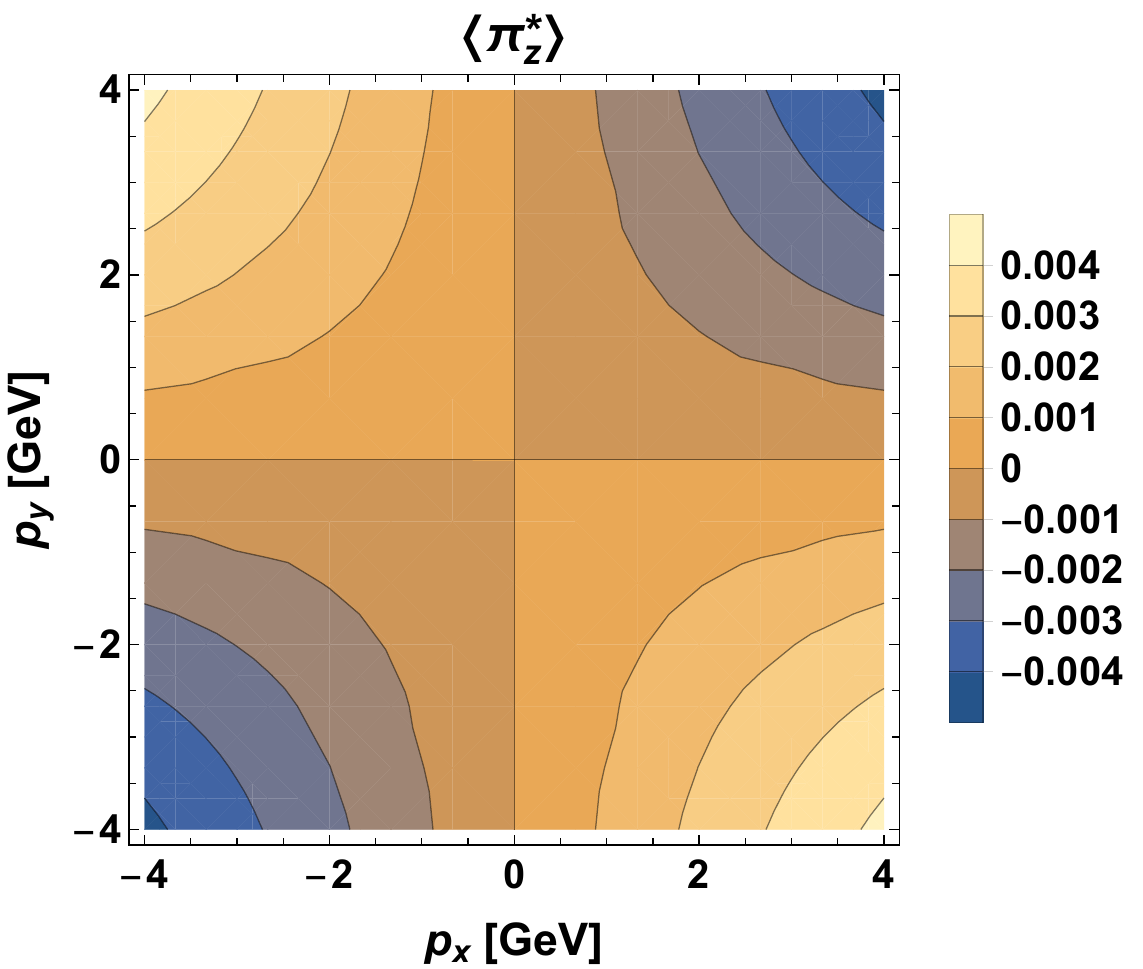}
		\subfigure[]{}\includegraphics[width=0.48\textwidth]{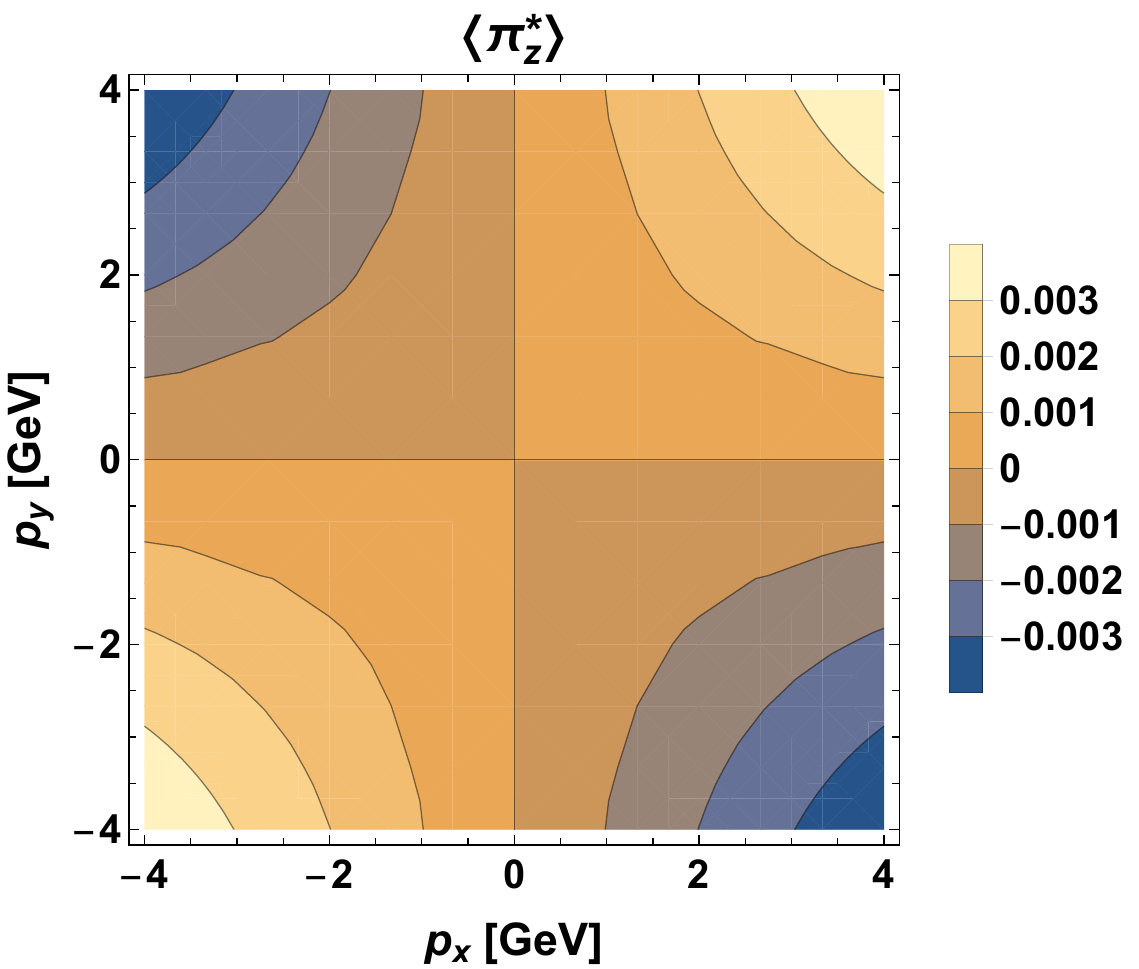}
		\caption{Same as Fig.~\ref{fig:polarization1} but for the centrality class $c$=30--60\%.} 
	 \label{fig:polarization3}
\end{figure*}

Our results presented in Figs. 1--3 (corresponding to the analyzed three centrality classes) show a quadrupole structure of the longitudinal polarization, whose sign depends on the choice of the definition of the spin polarization tensor $\omega_{\mu\nu}$. In the case where the spin polarization tensor is equal to the thermal vorticity, panels (a), we obtain an opposite sign compared to that found in the experiment. We note that a similar discrepancy was obtained in the earlier hydrodynamic calculations which used the relation $\omega_{\mu\nu} = \varpi_{\mu\nu}$. On the contrary, the use of the projected thermal vorticity in the definition of the spin polarization leads to the correct sign of the quadrupole structure, see panels (b). It is also interesting to note that the magnitude of the effect for the centrality class $c$=30--60\% is similar to the observed one (see Fig. 6 in ~\cite{Niida:2018hfw}, where the results for the centrality class $c$=10--60\% are shown). 

Our choice to use the projected thermal vorticity as a source of the spin polarization was motivated by the non-relativistic calculations which, in the natural way, use only the spatial components of the rotation $\partial_i v_j - \partial_j v_i$. Why this choice should be suitable for the description of the data remains an open question, which could perhaps be addressed if dynamical approaches describing the spin polarization in heavy-ion collisions became available.

Finally, we have showed how the hydrodynamic equations of motion can be used to build hydrodynamically consistent gradients of fluid velocity and temperature from the information just on the freeze-out surface. The general formulae summarized in the Appendix~\ref{sec:scg} can be used to calculate gradients terms like vorticity or shear-stress tensor from a freeze-out surface parametrization.

 \bigskip 

Also, we would like to note here that during the final preparation of this manuscript another paper following the idea of spin polarization being given by a  definition of vorticity different from thermal vorticity appeared \cite{Wu:2019eyi}. In fact the conclusions of this study show that the correct sign of the longitudinal spin polarization may be obtained using the so-called $T$-vorticity \cite{Becattini:2015ska}. One should stress, however, that the study was performed within a quite different dynamical model (3+1D viscous hydrodynamics with Glauber or AMPT initial conditions) from the one considered herein.

\begin{acknowledgments}
We thank Sergei Voloshin for illuminating discussions. This work was supported in part by the Polish National Science Center Grants   No. 2016/23/B/ST2/00717 and No. 2018/30/E/ST2/00432, and by the German Research Foundation (DFG) Collaborative Research Centre “SFB 1225
(ISOQUANT)” (A.M.). A.M. thanks Institute of Nuclear Physics PAN for the hospitality during the short term visit.
\end{acknowledgments}

\appendix

\section{Hydrodynamic equations for baryon-free matter\label{sec:hydmub0}}

At the top RHIC energies studied in this work, one can neglect the effects of baryon number density and use the hydrodynamic equations in a simplified form where the temperature $T$ and the flow four-vector $u^\mu$ are the only independent hydrodynamic variables. In this case, the equation of state can be encoded in the temperature dependence of the sound velocity, $c_s^2(T) = dp(T)/de(T)$, where $e$ is the energy density and $p$ is the pressure. 
A more general case is considered in Appendix~\ref{sec:scg}.

Herein we consider a perfect fluid characterized by the energy-momentum tensor
$T^{\mu\nu}=(e+p) u^{\mu}u^{\nu}- p g^{\mu\nu}$ (with $g^{00}=+1$). 
The energy and momentum conservation laws are expressed by the formula
\begin{equation}
\partial_\mu T^{\mu\nu}(x)=0.
\label{eq:enmomcon}
\end{equation}
The four equations contained in Eq.~(\ref{eq:enmomcon}) can be rewritten as a pair of the following two equations~\cite{Florkowski:2010zz}:
\begin{align}
    D u^{\alpha }&=\frac{1}{T}\nabla^{\alpha }T
\label{eq:A2},\\
    DT&=-T c_s^2 \partial_{\alpha}u^{\alpha},
 \label{eq:A3}
  \end{align}
  where we have defined $D=u^{\alpha}\partial_{\alpha}$ and $\nabla ^{\alpha }=\partial ^{\alpha }-u^{\alpha }D$ and used that $e+p=sT$ and $s(T)=dp(T)/dT$. Note that only three equations in (\ref{eq:A2}) are linearly independent. After simple manipulations, using Eqs.~(\ref{eq:A2}) and (\ref{eq:A3}) yield the formula for the temperature gradient 
\beq
\partial^{\alpha }T=T \left(D u^{\alpha }- c_s^2 u^{\alpha }\partial_\mu u^{\mu }\right).
\label{eq:tempgrad}
\eeq

This formula used in the second term on the right-hand side of Eq.~(\ref{eq:thvor}) gives $-1/(2T) (u_\mu D u_\nu - u_\nu D u_\mu)$. The first term on the right-hand side of Eq.~(\ref{eq:thvor}) is equal to $-1/(2T) [(u_\mu D u_\nu - u_\nu D u_\mu) + (\nabla_\mu u_\nu - \nabla_\nu u_\mu) ]$. For our form of the flow the term $\nabla_\mu u_\nu - \nabla_\nu u_\mu$ vanishes (this can be checked by a direct calculation), hence the contributions from the two terms in Eq.~(\ref{eq:thvor}) are equal.

%

\section{Hydrodynamically consistent gradients on the freeze-out surface\label{sec:scg}}

In general the freeze-out hyper-surface $\Sigma$ specifies the temperature $T(x)$, chemical potential $\mu(x)$ and velocity
$u^\mu(x)$ on the three dimensional subspace of the Minkowski space-time.  Using the projector operator $\tilde{\Delta}^{\mu\nu} = \left( g^{\mu\nu}-\frac{n^\mu n^\nu}{n\cdot n} \right) $, where $n^\mu$ is a non-null surface norm $n^\mu n_\mu=\pm1$
\footnote{Formulas in this section are the same for both mostly positive and mostly negative metric conventions.}
, we can define in-plane and out-of-plane projections of the covariant derivative on the freeze-out surface
%
\begin{align}
 \tilde \nabla_\mu \equiv  \tilde{\Delta}_\mu^{\phantom{\mu}\nu}D_\nu, \quad  \tilde D \equiv n^\nu D_\nu .
\end{align}
At each point on the freeze-out surface the full gradients of hydrodynamic fields can be decomposed to a sum
of in-plane and out-of-plane contributions
\begin{align}
  \begin{split}
  D_\mu T &= \frac{n_\mu}{n\cdot n} \tilde D T + \tilde \nabla_\mu T,\\
  D_\mu u_\nu &= \frac{n_\mu}{n\cdot n} \tilde  D u_\nu+\tilde \nabla_\mu u_\nu.
\end{split}\label{eq:Tugradients}
\end{align}
Without further assumptions about the fields on the surface $\Sigma$, only the in-plane gradients  can be calculated from the information on the surface. However, for the expansion governed by the energy-momentum conservation, the in- and out-of-plane gradients are related by the equations of motions.
For the case of a perfect fluid these equations reduce to
 \begin{align}
   u^\nu D_\nu e&= -(e + p) D_\sigma u^\sigma,\label{eq:econs} \\
 \frac{u^\nu}{u\cdot u} D_\nu u^{\mu} &=  \frac{ \Delta^{\mu\sigma} D_\sigma p }{e + p},
 \end{align}
 where the projection orthogonal to fluid velocity $u^\mu$ is given by $\Delta^{\mu\nu}\equiv g^{\mu\nu}-\frac{u^\mu u^\nu}{u\cdot u}$. Note that in a perfect fluid approximation, the entropy per baryon is conserved $u^\mu D_\mu (s/n_B)=0$, therefore the pressure gradient can be related to energy gradients using the sound velocity~\cite{Florkowski:2010zz}
\begin{equation}
  u^\mu D_\mu p(e,s/n_B) = c_s^2 u^\nu D_\nu e,\label{eq:udiv}.
  \end{equation}
Here we can use one of the following expressions
 \begin{eqnarray}
  c_s^2 &=& 
  \left( \frac{\partial P}{\partial e} \right)_{\!s/n_B} = \frac{n_B}{e+p} 
  \left( \frac{\partial P}{\partial n_B} \right)_{\!s/n_B} \nonumber \\
  &=&  \frac{n_B}{w} 
  \left( \frac{\partial w}{\partial n_B} \right)_{\!s/n_B},
  \label{cs2}
\end{eqnarray}
where $w = (e+P)/n_B$ is the specific enthalpy, see \cite{Misner:1974qy,Florkowski:2010zz}.

For the freeze-out surface $\Sigma$ with constant temperature and baryon chemical potential, the energy density and pressure are also constant along the freeze-out surface. We can drop the in-plane gradients of energy and pressure, and also use $\left.u^\nu D_\nu p\right|_\Sigma =  \frac{n\cdot u}{n \cdot n} \tilde D p  $  and $ \left.u^\nu D_\nu e\right|_\Sigma =  \frac{n\cdot u}{n \cdot n} \tilde D e  $ together with \rf{eq:udiv} to relate the out-of-plane gradients of energy and pressure,
\begin{equation}
      \tilde D p = c_s^2 \tilde D e .
\end{equation}
    
Eliminating energy from the equations of motion and collecting the in-plane and out-of-plane gradients we arrive at
\begin{align}
 &\left( \frac{u\cdot n }{n\cdot n u \cdot u}g^{\mu\sigma} +c_s^2\frac{\Delta^{\mu \rho}n_\rho n^\sigma}{u\cdot n n\cdot n}  \right) \tilde D u_\sigma=\nonumber\\
 &\quad\quad\qquad- \left(\frac{u^\kappa g^{\mu \lambda}}{u\cdot u} +c_s^2 \frac{\Delta^{\mu\rho}n_\rho g^{\kappa \lambda}}{u\cdot n} \right) \tilde \nabla_\kappa  u_\lambda.\label{eq:grad}
\end{align}
Note that the system of equations in \refeq{eq:grad} has only three independent components, because of the orthogonality to fluid velocity $u^\mu$, hence, it can be inverted if we restrict ourselves to the sub-space projected by $\Delta^{\mu\nu}$. Explicitly the necessary inverse is
\begin{align}
& \left( \frac{u\cdot n}{n\cdot n u\cdot u} \Delta^{\mu\nu}+c_s^2 \frac{\Delta^{\mu\rho}n_\rho n_\sigma \Delta^{\sigma \nu} }{u\cdot n n\cdot n} \right)^{-1}\label{eq:inv}
=\\
& \frac{n \cdot n u\cdot u}{u\cdot n}  \Delta^{\mu\nu}-c_s^2\frac{n \cdot n}{u\cdot n}\frac{  \Delta^{\mu\rho}n_\rho n_\sigma \Delta^{\sigma \nu}}{(u\cdot n)^2 + (u\cdot u) c_s^2 n_\sigma  \Delta^{\sigma \rho}n_\rho}. \nonumber
\end{align}
Then, applying \refeq{eq:inv} to the right hand side of \refeq{eq:grad} and performing straightforward simplifications, the out-of-plane velocity gradients are given by
\begin{align}
  \label{eq:unew}  \tilde D u_\nu &= - \frac{n\cdot n }{u\cdot n } \Big(  u_\kappa\Delta_{\nu \lambda}+\\
                 &c_s^2\frac{- n^\sigma \Delta_{\sigma \lambda} u_\kappa+ u\cdot n  \Delta_{\kappa \lambda}}{u\cdot u(u\cdot n)^2 +  c_s^2 n_\sigma  \Delta^{\sigma \rho}n_\rho}  \Delta_{\nu\rho}n^\rho\Big) \tilde \nabla^\kappa u^\lambda\nonumber
.\end{align}
\refeq{eq:unew} can be tested using known analytical solutions of hydrodynamic equations of motion, e.g. Gubser flow~\cite{Gubser:2010ze}.
The out-of-plane temperature gradients can be calculated from \refeq{eq:econs} and are given by the divergence of fluid velocity. Leaving the details of derivation to Appendix~\ref{sec:derivation} we quote the final result 
 \begin{align}
  \label{eq:Tnew}  \frac{ \tilde  D T}{T} &= - \frac{n\cdot n}{u\cdot n } c_s^2 D_\sigma u^\sigma  \frac{1 + \frac{n_B \mu}{sT}}{1+\frac{n_B}{s}\left( \frac{\partial \mu}{\partial T} \right)_{\!s/n_B}},
 \end{align}
 where $D_\sigma u^{\sigma}= n_\mu\tilde Du^{\mu} / n\cdot n+\tilde{\nabla}_\mu u^{\mu}$.
 
Note that thermodynamic quantities such as the speed of sound $c_s^2$ or the derivative of the baryon chemical potential are constant on the considered freeze-out surface and must be determined from the microscopic description of the fluid at freeze-out.

In summary, Eqs.~\eqref{eq:unew} and \eqref{eq:Tnew} give a simple prescription for constructing out-of-plane
gradients for fluid fields parametrized on an arbitrary non-null freeze-out surface of constant temperature and chemical potential. 

\section{Useful thermodynamic identities \label{sec:derivation}}

Starting from the conservation of energy given by Eq.~(\ref{eq:econs}), where we consider energy density to be a function of $T$ and $s/n_B$,
\begin{align}
      u^\nu D_\nu e(T, s/n_B)&= -(e + p) D_\sigma u^\sigma,
\end{align}
and using the fact that entropy per baryon is conserved, we obtain
\begin{align}
      \left(\frac{\partial e}{\partial T} \right)_{\!s/n_B}  \frac{u\cdot n}{n\cdot n} \, \tilde D T&= -(e + p) D_\sigma u^\sigma.
\end{align}
Here we used $D_\mu T = \frac{n_\mu}{n\cdot n} \tilde D T + \tilde{\nabla}_\mu T$ and dropped in-plane gradients. This equation directly gives the out-of-plane temperature gradient
\begin{align}
    \frac{ \tilde  D T}{T} &= - \frac{n\cdot n \, (e + p)}{u\cdot n \, T \left( \frac{\partial e}{\partial T} \right)_{\!s/n_B}}  D_\sigma u^\sigma.
    \label{tempgr}
\end{align}
Below we consider the thermodynamic quantity
\begin{equation}
\kappa = \frac{e+p}{T \left( \frac{\partial e}{\partial T} \right)_{\!s/n_B}} ,
\end{equation}
that enters Eq.~(\ref{tempgr}) and show that for small baryon number densities it is reduced to $c_s^2$. We first use the thermodynamic identity~\cite{Misner:1974qy,Florkowski:2010zz}
\begin{equation}
    e+p =  n_B \left( \frac{\partial e}{\partial n_B} \right)_{\!s/n_B} 
\end{equation}
and obtain
\begin{align}
\kappa 
    = 
    \frac{n_B}{T} 
    \frac{\left( \frac{\partial e}{\partial n_B} \right)_{\!s/n_B}}{\left( \frac{\partial e}{\partial T} \right)_{\!s/n_B}}
    =
     \frac{n_B}{T} 
     \left( \frac{\partial T}{\partial n_B} \right)_{\!s/n_B}.
   \end{align}
In the next step we switch to the specific enthalpy and connect it with the sound velocity through Eq.~(\ref{cs2}).  In this way we get
\begin{equation}
\!\!\!\!\kappa = 
     \frac{n_B}{T} \left( \frac{\partial T}{\partial w} \right)_{\!s/n_B}
     \left( \frac{\partial w}{\partial n} \right)_{\!s/n_B}  \!\!\!=
      c_s^2 \, \frac{w}{T} \left( \frac{\partial T}{\partial w} \right)_{\!s/n_B}\!\!\!.
\end{equation}
Since $e+p=sT+n_B\mu$, we may use $ w = (s/n_B) T + \mu$ to obtain
\begin{equation}
  \kappa =   c_s^2 \left(1+\frac{n_B\mu}{sT}\right) \left(1+ \frac{n_B}{s}  \left( \frac{\partial \mu}{\partial T} \right)_{\!s/n_B}\right)^{-1} 
\end{equation}
Thus, for small baryon densities, $n_B/s \ll 1$, or $\mu$ linear in $T$, we find that $\kappa \approx c_s^2$.


\bibliography{pv_ref}{}

\providecommand{\href}[2]{#2}\begingroup\raggedright\begin{thebibliography}{10}

\bibitem{Voloshin:2017kqp}
S.~A. Voloshin, ``{Vorticity and particle polarization in heavy ion collisions
  (experimental perspective)},'' \href{http://arxiv.org/abs/1710.08934}{{\tt
  arXiv:1710.08934 [nucl-ex]}}.
[EPJ Web Conf.17,10700(2018)].

\bibitem{Voloshin:2004ha}
S.~A. Voloshin, ``{Polarized secondary particles in unpolarized high energy
  hadron-hadron collisions?},''
\href{http://arxiv.org/abs/nucl-th/0410089}{{\tt arXiv:nucl-th/0410089
  [nucl-th]}}.

\bibitem{STAR:2017ckg}
{\bf STAR} Collaboration, L.~Adamczyk {\em et al.}, ``{Global $\Lambda$ hyperon
  polarization in nuclear collisions: evidence for the most vortical fluid},''
  \href{http://dx.doi.org/10.1038/nature23004}{{\em Nature} {\bf 548} (2017)
  62--65},
\href{http://arxiv.org/abs/1701.06657}{{\tt arXiv:1701.06657 [nucl-ex]}}.

\bibitem{Adam:2018ivw}
{\bf STAR} Collaboration, J.~Adam {\em et al.}, ``{Global polarization of
  $\Lambda$ hyperons in Au+Au collisions at $\sqrt{s_{_{NN}}}$ = 200 GeV},''
  \href{http://dx.doi.org/10.1103/PhysRevC.98.014910}{{\em Phys. Rev.} {\bf
  C98} (2018)  014910},
\href{http://arxiv.org/abs/1805.04400}{{\tt arXiv:1805.04400 [nucl-ex]}}.

\bibitem{dehaas:1915}
A.~Einstein and W.~de~Haas, ``{Experimenteller Nachweis der Ampereschen
  Molekularstroeme},'' {\em Deutsche Physikalische Gesellschaft, Verhandlungen}
  {\bf 17} (1915)  152.

\bibitem{RevModPhys.7.129}
S.~J. Barnett, \href{http://dx.doi.org/10.1103/RevModPhys.7.129}{``Gyromagnetic
  and electron-inertia effects,''{\em Rev. Mod. Phys.} {\bf 7} (Apr, 1935)
  129--166}. \url{https://link.aps.org/doi/10.1103/RevModPhys.7.129}.

\bibitem{Karpenko:2016jyx}
I.~Karpenko and F.~Becattini, ``{Study of $\Lambda $ polarization in
  relativistic nuclear collisions at $\sqrt{s_\mathrm {NN}}=7.7$ –200 GeV},''
  \href{http://dx.doi.org/10.1140/epjc/s10052-017-4765-1}{{\em Eur. Phys. J.}
  {\bf C77} (2017) no.~4, 213},
\href{http://arxiv.org/abs/1610.04717}{{\tt arXiv:1610.04717 [nucl-th]}}.

\bibitem{Li:2017slc}
H.~Li, L.-G. Pang, Q.~Wang, and X.-L. Xia, ``{Global $\Lambda$ polarization in
  heavy-ion collisions from a transport model},''
  \href{http://dx.doi.org/10.1103/PhysRevC.96.054908}{{\em Phys. Rev.} {\bf
  C96} (2017) no.~5, 054908},
\href{http://arxiv.org/abs/1704.01507}{{\tt arXiv:1704.01507 [nucl-th]}}.

\bibitem{Becattini:2007nd}
F.~Becattini and F.~Piccinini, ``{The Ideal relativistic spinning gas:
  Polarization and spectra},''
  \href{http://dx.doi.org/10.1016/j.aop.2008.01.001}{{\em Annals Phys.} {\bf
  323} (2008)  2452--2473},
\href{http://arxiv.org/abs/0710.5694}{{\tt arXiv:0710.5694 [nucl-th]}}.

\bibitem{Becattini:2009wh}
F.~Becattini and L.~Tinti, ``{The Ideal relativistic rotating gas as a perfect
  fluid with spin},'' \href{http://dx.doi.org/10.1016/j.aop.2010.03.007}{{\em
  Annals Phys.} {\bf 325} (2010)  1566--1594},
\href{http://arxiv.org/abs/0911.0864}{{\tt arXiv:0911.0864 [gr-qc]}}.

\bibitem{Becattini:2017gcx}
F.~Becattini and I.~Karpenko, ``{Collective Longitudinal Polarization in
  Relativistic Heavy-Ion Collisions at Very High Energy},''
  \href{http://dx.doi.org/10.1103/PhysRevLett.120.012302}{{\em Phys. Rev.
  Lett.} {\bf 120} (2018) no.~1, 012302},
\href{http://arxiv.org/abs/1707.07984}{{\tt arXiv:1707.07984 [nucl-th]}}.

\bibitem{Niida:2018hfw}
{\bf STAR} Collaboration, T.~Niida, ``{Global and local polarization of
  $\Lambda$ hyperons in Au+Au collisions at 200 GeV from STAR},''
  \href{http://dx.doi.org/10.1016/j.nuclphysa.2018.08.034}{{\em Nucl. Phys.}
  {\bf A982} (2019)  511--514},
\href{http://arxiv.org/abs/1808.10482}{{\tt arXiv:1808.10482 [nucl-ex]}}.

\bibitem{Becattini:2013vja}
F.~Becattini, L.~Csernai, and D.~J. Wang, ``{$\Lambda$ polarization in
  peripheral heavy ion collisions},''
  \href{http://dx.doi.org/10.1103/PhysRevC.93.069901,
  10.1103/PhysRevC.88.034905}{{\em Phys. Rev.} {\bf C88} (2013) no.~3, 034905},
  \href{http://arxiv.org/abs/1304.4427}{{\tt arXiv:1304.4427 [nucl-th]}}.
[Erratum: Phys. Rev.C93,no.6,069901(2016)].

\bibitem{Becattini:2016gvu}
F.~Becattini, I.~Karpenko, M.~Lisa, I.~Upsal, and S.~Voloshin, ``{Global
  hyperon polarization at local thermodynamic equilibrium with vorticity,
  magnetic field and feed-down},''
  \href{http://dx.doi.org/10.1103/PhysRevC.95.054902}{{\em Phys. Rev.} {\bf
  C95} (2017) no.~5, 054902},
\href{http://arxiv.org/abs/1610.02506}{{\tt arXiv:1610.02506 [nucl-th]}}.

\bibitem{Xie:2015xpa}
Y.~Xie, R.~C. Glastad, and L.~P. Csernai, ``{$\Lambda$ polarization in an exact
  rotating and expanding fluid dynamical model for peripheral heavy ion
  reactions},'' \href{http://dx.doi.org/10.1103/PhysRevC.92.064901}{{\em Phys.
  Rev.} {\bf C92} (2015) no.~6, 064901},
\href{http://arxiv.org/abs/1505.07221}{{\tt arXiv:1505.07221 [nucl-th]}}.

\bibitem{Boldizsar:2018akg}
B.~Boldizsar, M.~I. Nagy, and M.~Csanad, ``{Polarized baryon production in
  heavy ion collisions: an analytic hydrodynamical study},''
\href{http://arxiv.org/abs/1812.05587}{{\tt arXiv:1812.05587 [hep-ph]}}.

\bibitem{Cleymans:1992zc}
J.~Cleymans and H.~Satz, ``{Thermal hadron production in high-energy heavy ion
  collisions},'' \href{http://dx.doi.org/10.1007/BF01555746}{{\em Z. Phys.}
  {\bf C57} (1993)  135--148},
\href{http://arxiv.org/abs/hep-ph/9207204}{{\tt arXiv:hep-ph/9207204
  [hep-ph]}}.

\bibitem{BraunMunzinger:2001ip}
P.~Braun-Munzinger, D.~Magestro, K.~Redlich, and J.~Stachel, ``{Hadron
  production in Au - Au collisions at RHIC},''
  \href{http://dx.doi.org/10.1016/S0370-2693(01)01069-3}{{\em Phys. Lett.} {\bf
  B518} (2001)  41--46},
\href{http://arxiv.org/abs/hep-ph/0105229}{{\tt arXiv:hep-ph/0105229
  [hep-ph]}}.

\bibitem{Florkowski:2001fp}
W.~Florkowski, W.~Broniowski, and M.~Michalec, ``{Thermal analysis of particle
  ratios and p(t) spectra at RHIC},'' {\em Acta Phys. Polon.} {\bf B33} (2002)
  761--769,
\href{http://arxiv.org/abs/nucl-th/0106009}{{\tt arXiv:nucl-th/0106009
  [nucl-th]}}.

\bibitem{Becattini:2005xt}
F.~Becattini, J.~Manninen, and M.~Gazdzicki, ``{Energy and system size
  dependence of chemical freeze-out in relativistic nuclear collisions},''
  \href{http://dx.doi.org/10.1103/PhysRevC.73.044905}{{\em Phys. Rev.} {\bf
  C73} (2006)  044905},
\href{http://arxiv.org/abs/hep-ph/0511092}{{\tt arXiv:hep-ph/0511092
  [hep-ph]}}.

\bibitem{Andronic:2017pug}
A.~Andronic, P.~Braun-Munzinger, K.~Redlich, and J.~Stachel, ``{Decoding the
  phase structure of QCD via particle production at high energy},''
  \href{http://dx.doi.org/10.1038/s41586-018-0491-6}{{\em Nature} {\bf 561}
  (2018) no.~7723, 321--330},
\href{http://arxiv.org/abs/1710.09425}{{\tt arXiv:1710.09425 [nucl-th]}}.

\bibitem{Broniowski:2001we}
W.~Broniowski and W.~Florkowski, ``{Explanation of the RHIC p(T) spectra in a
  thermal model with expansion},''
  \href{http://dx.doi.org/10.1103/PhysRevLett.87.272302}{{\em Phys. Rev. Lett.}
  {\bf 87} (2001)  272302},
\href{http://arxiv.org/abs/nucl-th/0106050}{{\tt arXiv:nucl-th/0106050
  [nucl-th]}}.

\bibitem{Florkowski:2017ruc}
W.~Florkowski, B.~Friman, A.~Jaiswal, and E.~Speranza, ``{Relativistic fluid
  dynamics with spin},''
  \href{http://dx.doi.org/10.1103/PhysRevC.97.041901}{{\em Phys. Rev.} {\bf
  C97} (2018) no.~4, 041901},
\href{http://arxiv.org/abs/1705.00587}{{\tt arXiv:1705.00587 [nucl-th]}}.

\bibitem{Florkowski:2017dyn}
W.~Florkowski, B.~Friman, A.~Jaiswal, R.~Ryblewski, and E.~Speranza,
  ``{Spin-dependent distribution functions for relativistic hydrodynamics of
  spin-1/2 particles},''
  \href{http://dx.doi.org/10.1103/PhysRevD.97.116017}{{\em Phys. Rev.} {\bf
  D97} (2018) no.~11, 116017},
\href{http://arxiv.org/abs/1712.07676}{{\tt arXiv:1712.07676 [nucl-th]}}.

\bibitem{Florkowski:2018myy}
W.~Florkowski, E.~Speranza, and F.~Becattini, ``{Perfect-fluid hydrodynamics
  with constant acceleration along the stream lines and spin polarization},''
  \href{http://dx.doi.org/10.5506/APhysPolB.49.1409}{{\em Acta Phys. Polon.}
  {\bf B49} (2018)  1409},
\href{http://arxiv.org/abs/1803.11098}{{\tt arXiv:1803.11098 [nucl-th]}}.

\bibitem{Florkowski:2018ahw}
W.~Florkowski, A.~Kumar, and R.~Ryblewski, ``{Thermodynamic versus kinetic
  approach to polarization-vorticity coupling},''
  \href{http://dx.doi.org/10.1103/PhysRevC.98.044906}{{\em Phys. Rev.} {\bf
  C98} (2018) no.~4, 044906},
\href{http://arxiv.org/abs/1806.02616}{{\tt arXiv:1806.02616 [hep-ph]}}.

\bibitem{Becattini:2018duy}
F.~Becattini, W.~Florkowski, and E.~Speranza, ``{Spin tensor and its role in
  non-equilibrium thermodynamics},''
  \href{http://dx.doi.org/10.1016/j.physletb.2018.12.016}{{\em Phys. Lett.}
  {\bf B789} (2019)  419--425},
\href{http://arxiv.org/abs/1807.10994}{{\tt arXiv:1807.10994 [hep-th]}}.

\bibitem{Florkowski:2018fap}
W.~Florkowski and R.~Ryblewski, ``{Hydrodynamics with spin --- pseudo-gauge
  transformations, semi-classical expansion, and Pauli-Lubanski vector},''
\href{http://arxiv.org/abs/1811.04409}{{\tt arXiv:1811.04409 [nucl-th]}}.

\bibitem{Sun:2018bjl}
Y.~Sun and C.~M. Ko, ``{Azimuthal angle dependence of the longitudinal spin
  polarization in relativistic heavy ion collisions},''
  \href{http://dx.doi.org/10.1103/PhysRevC.99.011903}{{\em Phys. Rev.} {\bf
  C99} (2019) no.~1, 011903},
\href{http://arxiv.org/abs/1810.10359}{{\tt arXiv:1810.10359 [nucl-th]}}.

\bibitem{Montenegro:2018bcf}
D.~Montenegro and G.~Torrieri, ``{Causality and dissipation in relativistic
  polarizeable fluids},''
\href{http://arxiv.org/abs/1807.02796}{{\tt arXiv:1807.02796 [hep-th]}}.

\bibitem{Weickgenannt:2019dks}
N.~Weickgenannt, X.-L. Sheng, E.~Speranza, Q.~Wang, and D.~H. Rischke,
  ``{Kinetic theory for massive spin-1/2 particles from the Wigner-function
  formalism},''
\href{http://arxiv.org/abs/1902.06513}{{\tt arXiv:1902.06513 [hep-ph]}}.

\bibitem{Hattori:2019lfp}
K.~Hattori, M.~Hongo, X.-G. Huang, M.~Matsuo, and H.~Taya, ``{Fate of spin
  polarization in a relativistic fluid: An entropy-current analysis},''
  \href{http://dx.doi.org/10.1016/j.physletb.2019.05.040}{{\em Phys. Lett.}
  {\bf B795} (2019)  100--106},
\href{http://arxiv.org/abs/1901.06615}{{\tt arXiv:1901.06615 [hep-th]}}.

\bibitem{Xie:2019jun}
Y.~Xie, D.~Wang, and L.~P. Csernai, ``{Fluid Dynamics Study of the $\Lambda$
  Polarization for Au+Au Collisions at $\sqrt{s_{NN}}=200$ GeV},''
\href{http://arxiv.org/abs/1907.00773}{{\tt arXiv:1907.00773 [hep-ph]}}.

\bibitem{Florkowski:2019qdp}
W.~Florkowski, A.~Kumar, R.~Ryblewski, and R.~Singh, ``{Spin polarization
  evolution in a boost invariant hydrodynamical background},''
\href{http://arxiv.org/abs/1901.09655}{{\tt arXiv:1901.09655 [hep-ph]}}.

\bibitem{Voloshin}
S.~Voloshin, ``{International Workshop XLVII on Gross Properties of Nuclei and
  Nuclear Excitations},''
\newblock Hirschegg, Kleinwalsertal, Austria, January 13-19, 2019.

\bibitem{Broniowski:2002wp}
W.~Broniowski, A.~Baran, and W.~Florkowski, ``{Thermal model at RHIC. Part 2.
  Elliptic flow and HBT radii},''
  \href{http://dx.doi.org/10.1063/1.1570571}{{\em AIP Conf. Proc.} {\bf 660}
  (2003) no.~1, 185--195},
\href{http://arxiv.org/abs/nucl-th/0212053}{{\tt arXiv:nucl-th/0212053
  [nucl-th]}}.

\bibitem{Karpenko:2018erl}
I.~Karpenko and F.~Becattini, ``{Lambda polarization in heavy ion collisions:
  from RHIC BES to LHC energies},''
  \href{http://dx.doi.org/10.1016/j.nuclphysa.2018.10.067}{{\em Nucl. Phys.}
  {\bf A982} (2019)  519--522},
\href{http://arxiv.org/abs/1811.00322}{{\tt arXiv:1811.00322 [nucl-th]}}.

\bibitem{Florkowski:2004du}
W.~Florkowski, W.~Broniowski, and A.~Baran, ``{Strange particle production in a
  single-freeze-out model},''
  \href{http://dx.doi.org/10.1088/0954-3899/31/6/064}{{\em J. Phys.} {\bf G31}
  (2005)  S1087--S1090},
\href{http://arxiv.org/abs/nucl-th/0412077}{{\tt arXiv:nucl-th/0412077
  [nucl-th]}}.

\bibitem{DeGroot:1980dk}
S.~R. De~Groot, {\em {Relativistic Kinetic Theory. Principles and
  Applications}}.
\newblock
1980.
\newblock

\bibitem{Leader:2001}
E.~Leader, {\em {Spin in Particle Physics}}.
\newblock {Cambridge}, 2001.

\bibitem{baran}
A.~Baran, ``{Description of azimuthal asymmetry in relativistic heavy-ion
  collisions based on a thermal model of particle production},''
\newblock PhD Thesis (W. Broniowski - supervisor), Institute of Nuclear
  Physics, Krakow, Poland, 2004.

\bibitem{Wu:2019eyi}
H.-Z. Wu, L.-G. Pang, X.-G. Huang, and Q.~Wang, ``{Local spin polarization in
  high energy heavy ion collisions},''
\href{http://arxiv.org/abs/1906.09385}{{\tt arXiv:1906.09385 [nucl-th]}}.

\bibitem{Becattini:2015ska}
F.~Becattini, G.~Inghirami, V.~Rolando, A.~Beraudo, L.~Del~Zanna, A.~De~Pace,
  M.~Nardi, G.~Pagliara, and V.~Chandra, ``{A study of vorticity formation in
  high energy nuclear collisions},''
  \href{http://dx.doi.org/10.1140/epjc/s10052-015-3624-1,
  10.1140/epjc/s10052-018-5810-4}{{\em Eur. Phys. J.} {\bf C75} (2015) no.~9,
  406}, \href{http://arxiv.org/abs/1501.04468}{{\tt arXiv:1501.04468
  [nucl-th]}}.
[Erratum: Eur. Phys. J.C78,no.5,354(2018)].

\bibitem{Florkowski:2010zz}
W.~Florkowski, {\em {Phenomenology of Ultra-Relativistic Heavy-Ion
  Collisions}}.
\newblock World Scientific, Singapore, 2010,
2010.
\newblock

\bibitem{Misner:1974qy}
C.~W. Misner, K.~S. Thorne, and J.~A. Wheeler, {\em {Gravitation}}.
\newblock W. H. Freeman, San Francisco,
1973.
\newblock

\bibitem{Gubser:2010ze}
S.~S. Gubser, ``{Symmetry constraints on generalizations of Bjorken flow},''
  \href{http://dx.doi.org/10.1103/PhysRevD.82.085027}{{\em Phys. Rev.} {\bf
  D82} (2010)  085027},
\href{http://arxiv.org/abs/1006.0006}{{\tt arXiv:1006.0006 [hep-th]}}.

\end{thebibliography}\endgroup
\bibliographystyle{utphys}

\end{document}